\documentclass[fleqn,10pt]{wlscirep}
\usepackage[utf8]{inputenc}
\usepackage[T1]{fontenc}
\usepackage{siunitx}
\title{Space-division multiplexed quantum key distribution exploiting multi-plane light conversion for few-mode fibers

}

\author[1,*,+]{Qian Zhang}
\author[1,2,+]{Stefan Krause}
\author[2]{Felix Kunzmann}
\author[1,3]{Juergen W Czarske}

\affil[1]{Laboratory of Measurement and Sensor Systems, Reinhart Koselleck Group, TU Dresden, Dresden, Germany}
\affil[2]{Fraunhofer IIS, Division Engineering of Adaptive Systems EAS, Dresden, Saxony, Germany}
\affil[3]{Faculty of Physics, School of Science, TU Dresden, Dresden, Germany
}

\affil[*]{qian.zhang@tu-dresden.de}

\affil[+]{these authors contributed equally to this work}

\begin{abstract}

As quantum key distribution (QKD) progresses from laboratory demonstrations toward practical deployment, quantum communication networks increasingly require higher key rates and the ability to distribute independent secret keys among multiple users and network nodes.
In this paper we present a promsing approach with multi-plane light conversion (MPLC) for demultiplexing entangled photons, transmitted through a few-mode fiber (FMF).
We experimentally demonstrated a spatially multiplexed BBM92 QKD scheme. The modes are selectively excited through separate single-mode-fibers and subsequently separated by MPLC into distinct output ports.
Unlike high-dimensional QKD based on coherent modal superpositions, our approach exploits distinguishable guided modes as parallel channels while preserving the entanglement required for QKD.
For the multiplexed links, we obtain quantum bit error rates of $1.9 \pm  0.4\%$ for the channel 1 and $6.8 \pm 0.8\%$ for the channel 2. 
These results are relevant for scalable quantum-secured networks, space-division-multiplexed QKD systems, multi-user entanglement distribution, and future quantum internet architectures, where parallel quantum channels must be implemented without compromising the quantum correlations required for security.

\end{abstract}
\begin{document}

\flushbottom
\maketitle

\thispagestyle{empty}
\noindent 

\section*{Introduction}

Modern communication networks support an ever-growing volume of data exchange, in which vast amounts of sensitive information are transmitted across highly interconnected infrastructures, making secure encryption and data protection increasingly essential ~\cite{rothe2023securing}.
In particular, rapid advances in quantum computing and quantum information technologies pose more and more a thread to conventional cryptographic systems whose security relies on computational complexity~\cite{pirandola2020advances}.
Quantum key distribution (QKD) has therefore emerged as a promising candidate for future secure communication, as it offers information-theoretic security rooted in the fundamental principles of quantum mechanics, such as the no-cloning theorem~\cite{gisin2002quantum,scarani2009security}. Any attack on the quantum channel by an eavesdropper inevitably introduces detectable disturbances into the transmitted quantum states, allowing the legitimate communicating parties to identify potential eavesdropping by monitoring the quantum bit error rate (QBER) and to distill a provably secure key.
\\ 
In recent years, QKD has gradually progressed from laboratory demonstrations to metropolitan links, backbone networks, and deployed fiber infrastructures~\cite{liu2023experimental,pathak2023phase,pittaluga2025long,liao2017satellite,li2025microsatellite,chen2021integrated}. However, its large-scale deployment remains constrained by the efficient utilization of channel resources and the complexity of system integration.
Most fiber-based QKD systems are implemented in single-mode fibers, where the available degrees of freedom are typically limited to polarization, phase, time bin, or wavelength~\cite{cozzolino2019high}. Although these degrees of freedom have enabled impressive progress, further scaling of quantum-secured communication networks requires new approaches that can increase the information capacity or parallelism of quantum channels without substantially increasing the footprint, loss, or alignment complexity of the system. 
Moreover, emerging quantum-network applications, including multi-user quantum communication and quantum internet architectures based on entanglement distribution and swapping, are expected to require the parallel transmission of multiple quantum channels~\cite{pant2019routing}.

Space-division multiplexing (SDM) provides a natural pathway for increasing the capacity of optical fiber links~\cite{richardson2013space}. In classical optical communications, multicore fibers (MCFs) and few-/multimode fibers (FMF/MMFs) have been widely recognized as important platforms for space-division multiplexing~\cite{shibahara2018dmd, qiao2024205}. Correspondingly, in quantum communications, exploiting spatial degrees of freedom to construct parallel quantum channels has attracted increasing attention~\cite{zia2024modal,vertesi2010closing}. Previous studies have shown that MCFs can provide a relatively stable physical platform for high-dimensional path-encoded QKD, ranging from early demonstrations of high-dimensional decoy-state QKD in MCFs to recent implementations of four-dimensional high-dimensional QKD over deployed MCF link~\cite{zahidy2024practical,dos2026high}. 
Related studies have also explored spatially structured fiber platforms, including air-core fiber~\cite{cozzolino2019orbital}/hollow-core fiber~\cite{habib2025hollow}, FMF~\cite{cao2020distribution} and
MCF~\cite{gomez2021multidimensional}, for high-dimensional quantum-state distribution and multidimensional quantum communication. These results demonstrate the feasibility of using spatially multiplexed fiber structures for QKD. However, most existing studies have focused on multicore path encoding~\cite{zahidy2024practical}or high-dimensional protocol implementations, while the use of different spatial modes in FMFs as independent quantum-channel resources remains underdeveloped.

In contrast to MCFs, where spatial channels are physically separated into distinct cores, FMFs support multiple guided eigenmodes within a single core, thereby enabling multiple spatial channels to be constructed in a compact single-fiber platform. If these modes can be selectively excited at the transmitter and efficiently separated at the receiver, FMFs could provide not only a mode-division multiplexing (MDM) medium for classical optical communications, but also a promising platform for parallel quantum-signal transmission~\cite{zia2024modal}. Nevertheless, the use of guided modes as quantum-channel resources is more challenging than multicore path encoding, because modal coupling and intermodal crosstalk are inherently stronger in FMFs. Consequently, selective mode excitation and low-crosstalk mode readout become critical prerequisites for extending FMF systems to quantum communication~\cite{zahidy2024practical}.

Multi-plane light conversion (MPLC) has emerged as a powerful approach for spatial-mode manipulation and mode sorting~\cite{fontaine2019laguerre}. By implementing tailored linear transformations between orthogonal spatial modes, MPLC can perform mode-selective conversion, multiplexing, and demultiplexing in a compact all-optical architecture. Recent studies have shown that MPLC can enable high-dimensional spatial-mode measurements for QKD and can also be applied to mode-division (de-)multiplexing in FMFs, providing an effective tool for separating quantum and classical signals in the mode domain ~\cite{lib2025high,gervaziev2020mode}. These capabilities make MPLC particularly attractive for spatial-mode quantum communication, where controlled mode excitation and low-crosstalk mode readout are essential. Therefore, integrating the modal transmission capability of FMFs with the mode-separation capability of MPLC offers a promising route toward space-division-multiplexed quantum links.

\begin{figure}[t]
    \centering
    \includegraphics[width=0.99\linewidth]{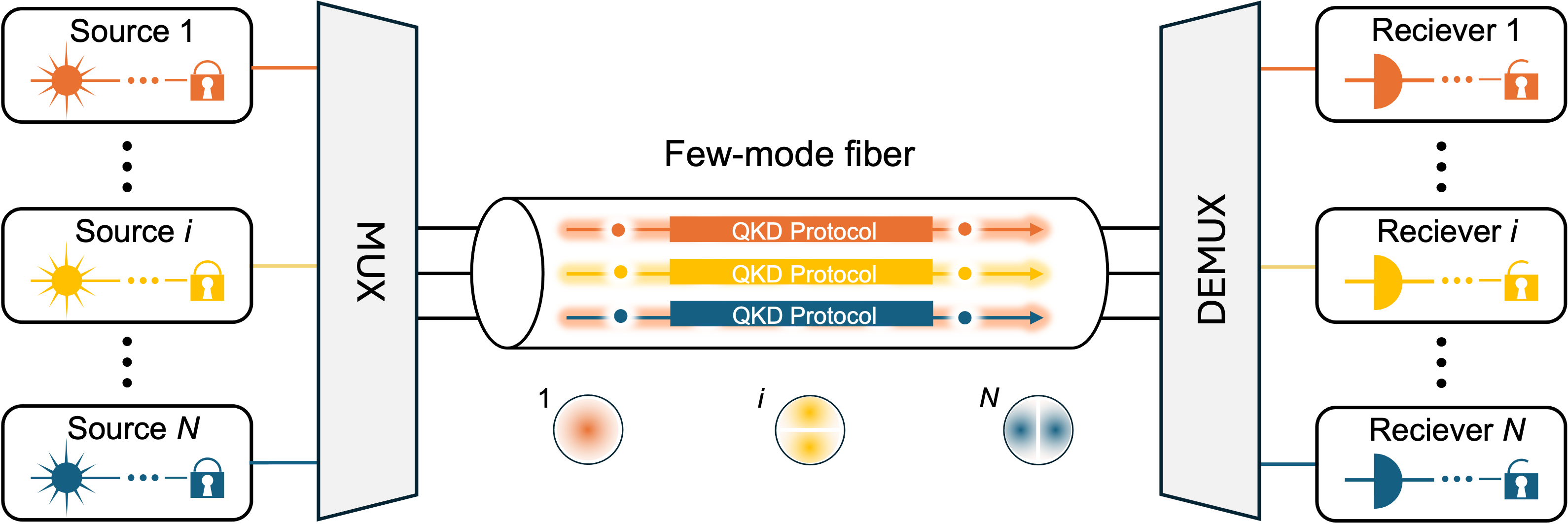}
    \caption{Proposed concept for parallel channels for QKD using FMF. The number of available quantum channels is determined by the number of spatial modes supported by the MPLC. In the present case, the implemented MPLC supports only a few spatial modes; however, the concept allows for scaling to a large number of quantum channels.}
    \label{fig:scheme}
\end{figure}

In this work, we propose and experimentally demonstrate a space-division-multiplexed BBM92~\cite{bennett1992quantum,waks2002security} QKD scheme using a FMF and an MPLC-based mode demultiplexer, as shown in Fig.~\ref{fig:scheme}. In our implementation, entangled photons are launched into a FMF through two single-mode-fiber inputs, where different guided modes are selectively excited by controlling the input coupling conditions. After transmission, the output modes are separated by an MPLC and directed to different detection channels, enabling distinct spatial modes to function as independent quantum channels. This approach differs from conventional high-dimensional spatial-mode QKD, which generally relies on coherent superpositions of multiple modes and measurements in multiple mutually unbiased bases. Instead, our work exploits the spatial modes of a FMF as separate transmission channels for entanglement-based QKD. We experimentally evaluate the mode-demultiplexing performance, two-photon visibility, and preservation of polarization entanglement.
We then implemented the BBM92 protocol over two spatial channels, achieving QBERs of 1.9\% and 6.8\% and secure key rates of 550 bit/s and 30 bit/s, respectively.
The experiments with MPLC have shown that independent channels in FMFs do not destroy the quantum correlations required for QKD. In conclusion, MPLC is very promising for spatially multiplexed fiber quantum communication.

\section*{Results}

\subsection*{Experimental setup}

\begin{figure}
    \centering
    \includegraphics[width=0.99\linewidth]{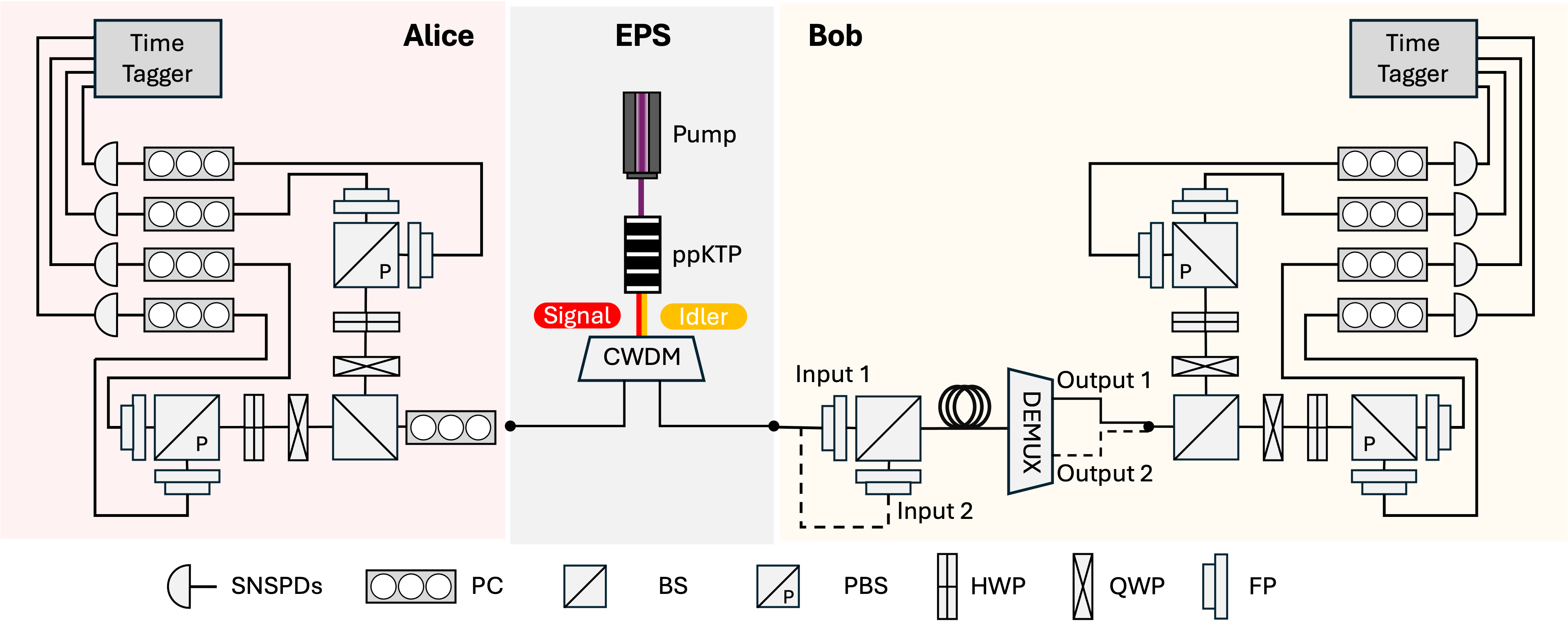}
    \caption{Scheme of the experimental setup. Abbreviations: BS, beam splitter; CWDM, coarse wavelength division multiplexing; EPS, entangled photon source; FP, fiber port; HWP, half wave plate; PBS, polarization beam splitter; PC, polarization controller; QWP, quarter-wave plate; SNSPD, superconducting nanowire single-photon detectors.}
    \label{fig:setup}
\end{figure}

The experimental setup is schematically shown in Fig.~\ref{fig:setup} and described in detail in \cite{nande2025tcep}. In brief, polarization-entangled photon pairs were generated through spontaneous parametric down-conversion (SPDC) in periodically poled potassium titanyl phosphate (PPKTP) crystals. A continuous-wave pump laser at a wavelength of $\lambda_p = 775$ nm (Toptica) was used to pump the PPKTP crystals in a beam displacer configuration \cite{hong2024polarization,fazili2024simple}, which were temperature-stabilized at $34 ^\circ\mathrm{C}$ to satisfy the quasi-phase-matching condition. The pump power before the crystal was set to 85 µW. After SPDC, residual pump photons were removed using a combination of dichroic mirrors and coarse wavelength division multiplexer (CWDM) which also separates the two non-degenerate photons from each other.
The central wavelengths of the signal and idler photon channels were $\lambda_s = 1530$ nm and
 $\lambda_i = 1570$ nm, respectively, with spectral bandwidths of $\Delta\lambda = 20$ nm. A fiber-based polarization controller (General Photonics) was used to compensate polarization transformations in the fiber via an iterativ approach. 
 
One photon of each entangled pair was transmitted through the proposed spatial channel of FMF, while the other photon served as the reference arm for coincidence measurements. At the transmitter side, the quantum signal was splitted into 2 channels and coupled into a FMF through two SMF inputs. By adjusting the input coupling position and angle, different guided modes of the FMF were selectively excited. The FMF (SI-FMF 4LP, Prysmian) had a core diameter of 15 $\mu\mathrm{m}$, a numerical aperture of $\mathrm{NA = 0.1}$, and a length of 10 m. 
This FMF supports 4 LP modes in the C-Band.
In this work, the quantum signals are multiplexde into $LP_{01}$ and $LP_{11b}$ as seprated channels. 

At the receiver side, 
to avoid quantum-state collapse induced by polarization-dependent optical components, we employ a polarization-independent diffractive optical device. The output field from the FMF was decomposed by a multi-plane light conversion (MPLC, Cailabs PROTEUS-S-6-SI-PRY) module for spatial-mode demultiplexing.
The insertion loss for $LP_{01}$ and $LP_{11b}$ is 2.0 dB and 2.3 dB, respectively.

For BBM92 measurement, both Alice and Bob receiver were equipped with polarization analysis modules consisting of a 50:50 beamsplitter cube for random bases choice, quarter- and half-wave plates for polarization adjustment and polarizing beam splitters in a free beam configuration. Input and output coupling was performed with single mode fiber couplers. Measurements were performed in two mutually unbiased polarization bases, namely the (H/V) basis and the (D/A) basis. The photons were detected by superconducting nanowire single-photon detectors (SNSPDs, IDQuantique) with detection efficiencies of about 85\% and dark count rates of about 10 cts/s. Detection events were recorded by a time-tagging module (qutools) with a jitter below 20 ps. 

Coincidence events were identified using a coincidence window of 100 ps after finding relative temporal offsets between all photon detector combinations of Alice and Bob via cross correlation analysis. From the measured time stamps the key material and statistical information were extracted after applying key sifting and error correction. QBER was calculated by transmitting parts of the key material from Bob to Alice for a direct comparison. The transmitted key material was afterwards discarded.\\ 

\subsection*{Experimental results}

Before performing the entangled-photon measurements, we first characterized the transmission and mode-demultiplexing properties of the system using a conventional laser at the same wavelength as the signal photons. This calibration allowed us to optimize the input coupling conditions and the MPLC alignment such that the optical power at the target output port was maximized. This step is particularly important in FMFs, where degenerate or near-degenerate modes have identical or very similar propagation constants and can therefore couple to each other during propagation. As a result, the experimentally observed output field may correspond to a superposition within the degenerate mode group rather than a single ideal eigenmode. We therefore optimized the system based on the power collected at the desired MPLC output channel, ensuring efficient excitation and readout of the target spatial channel before switching to entangled-photon measurements. 
After calibration, the insertion loss of home-built MUX has a value of 3.24 dB for Input 1 and 7.3 dB for Input 2. The transmission efficiencies from Input 1 to Output 1 and from Input 2 to Output 2 reached approximately 15\% and 5.5\%, corresponding to total  losses of 8.24 dB and 12.6 dB, respectively.  We would like to emphasize that especially the input coupling contributes to these high damping values, where 3 dB for each channel arise only through the beam splitter cube at the FMF input. These losses will be tackled in the near future.
The measured crosstalk values were -8.53 dB for the Input 1–Output 2 leakage relative to the Input 1–Output 1 channel, and -4.70 dB for the Input 2–Output 1 leakage relative to the Input 2–Output 2 channel.


\begin{figure}[tbp]
    \centering
    \includegraphics[width=0.99\linewidth]{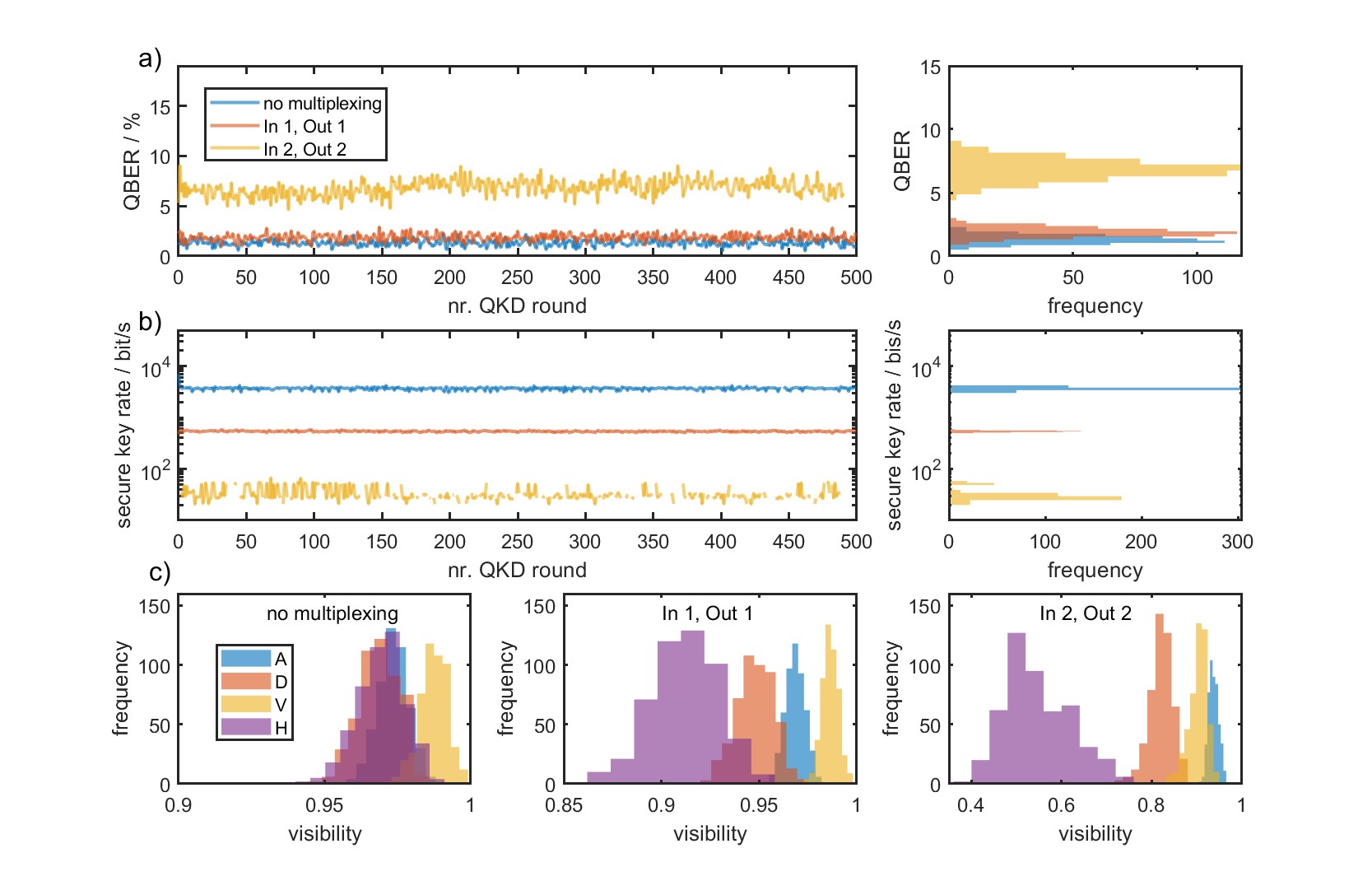}
    \caption{Results of the BBM92 QKD experiment with (red and yellos) and without spatial multiplexing. QBER (a), secure key rate (b) and average visibility (c) as a function of QKD round together with the 
    corresponding histograms. For the three test cases, the QKD protocol was continuously operated for 40 min in the non-multiplexed reference configuration, 115 min for the Input 1–Output 1 channel, and 184 min for the Input 2–Output 2 channel.}
    \label{fig:results}
\end{figure}


In this proof-of-concept experiment, our aim is to demonstrate that, when polarization entangled photons are multiplexed in different modes and transferred through a FMF, the entanglement fidelity is maintained since this multiplexing approach does not exclude or prefer any polarization state. This is a prerequisite, not only for the entanglement based BBM92 protocol, but also for single photon polarization based protocols such as BB84. 

\begin{figure}[htb]
    \centering
    \includegraphics[width=0.7\linewidth]{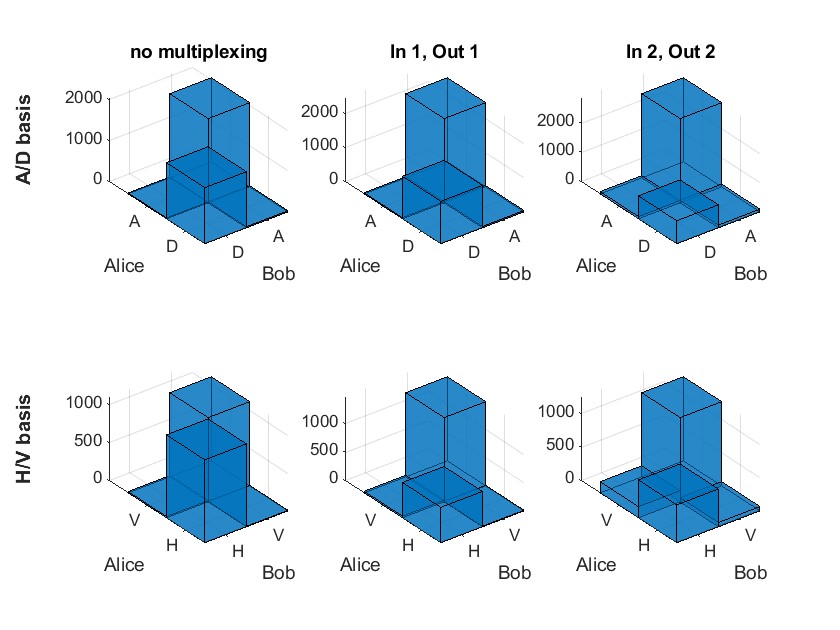}
    \caption{Average coincidence counts per QKD round for the QKD link without and with (In 1, Out 1 and In 2 Out 2) multiplexing for the two bases A/D and H/V.}
    \label{fig:coincidences}
\end{figure}

The experimental results of QKD with and without spatial multiplexing are shown in Fig.~\ref{fig:results}(a). 
It is clear from Fig.~\ref{fig:results}(a) that the QBER remains below the 11\% security threshold~\cite{gisin2002quantum} for all three configurations: the non-multiplexed reference case, the Input 1–Output 1 channel, and the Input 2–Output 2 channel.
While utizlizing Input 1 and Output 1, QBER increases only slightly from $1.5 \pm 0.4$\% (no multiplexing) to $1.9 \pm 0.4$\% (multiplexing with input 1 and output 1). However, when changing to the multiplexing channel Input 2 and Output 2, a singificant increase of QBER to $6.8 \pm 0.8$\% is observed. Nontheless, the low standard deviation of QBER averaged over about 500 QKD rounds illustrates the long term stability of the established, multiplexed QKD link. Missing QBER values, especially for the QKD link input 2 and output 2, result from a missmatch between estimated and real QBER and a subsequently failed error correction. Secure key rates ranging on average from 550 bit/s to 30 bit/s were achieved for the Input 1–Output 1 and Input 2–Output 2 channels.\\
Regarding the achieved secure key rate, shown in Fig.~\ref{fig:results}(b), it is obvious that the implemented multiplexing scheme introduces currently too much losses and can thus not outperform the secure key rate achieved without multiplexing. This is, however, not the scope of this paper. For improved secure key rate we propose increasing the coupling efficiency of the multiplexing especially at the input of the FMF which is currently the limiting factor. In this work, we used a non-polarizing beam splitter cube and two orthogonally aligend fiber couplers to couple the photons at two discrete angles into the FMF. While the beam splitter cube already introduces 50\% loss to each channel, it is in addition very likely, that imperfections of the beam splitter cube also result in polarization transformations which can not be compensated and reduce QBER. This is underlined when comparing the measured visibilities in Fig.~\ref{fig:results}(c) which are given by the coincidence ratio of the difference and sum of maximum and minimum coincidence counts $C$ for a given polarization state, e.g. $H$, according to $V_H = ({C_H-C_V})/({C_H+C_V})$. For multiplexing with Input 2 and Output 2, the visibility is mainly reduced vor coincidences appearing in H. Damping, which can also affect the visibility since it lowers the ratio of coincidences to accidentials would, however, affect A, D, V and H equally. This indicates a polarization effect being responsible for the visibility decrease.\\
A similar effect is noticable when inspecting the averaged coincidences for the discrete polarization states A, D, V and H at Alice and Bob depicted in Fig.~\ref{fig:coincidences}. For QKD without multiplexing the coincidence values reach comparable valules. Only the number of detected coincidences for A are increased by a factor of about 1.5 in comparison to D. However, in the multiplexing scenario this difference is further increasd to a factor of about 2 (In 1, Out 1) and 3.5 (In 2, Out 2) for A and D and a factor of about 2.3 (both modes) for H and V. This further suggests at least some polarization dependent influence on the multiplexed photons which is again very likely introduced at the FMF input.

\section*{Conclusion}

In this paper we have demonstrated distribution of polarization entangled photons over a mode multiplexing FMF device maintaining the entanglement fidelity required for BBM92 QKD. 
The BBM92 protocol is an entanglement-based QKD scheme in which Alice and Bob perform local measurements on their respective photons in two complementary polarization bases. It does not involve Bell-state measurements~\cite{bennett1992quantum}. Potential eavesdropping is assessed statistically by monitoring the correlations and quantum bit error rates in the two measurement bases, which allows Alice and Bob to bound Eve’s information and determine whether a secure key can be extracted. However, the crucial factor is the increased secure key rate enabled by the channels within the FMF. This promising approach has already been demonstrated. Nevertheless, low coupling efficiencies currently limit the transmission rate. In the future, however, this approach will enable significantly higher secure key rates with reduced losses.
Since the key rate scales linearly with the utilized number of sources and receivers which can be connected via a FMF using several spatial modes in parallel higher key rates are possible. This distribution modality is, on the other hand, not accessible via single mode fibers.  
The demonstrated channels were obtained by optimizing the input coupling and using a static MPLC demultiplexer, so the performance relies on the temporal stability of the FMF transmission. Future implementations could improve robustness either by using weakly coupled or ultra-low-crosstalk FMFs~\cite{ge20196} to suppress intermodal coupling, or by incorporating adaptive optics~\cite{krause2026quantum,rothe2023securing}, wavefront-shaping feedback~\cite{zhang2022learning,rothe2025unlocking,sun2022quantitative}, or dynamically reconfigurable MPLC~\cite{a2025self} to compensate time-varying modal crosstalk. 
Moreover, advanced multiplexing and demultiplexing techniques can be further enhanced by artificial intelligence~\cite{czarske2026complex,Huang2026HybridBrightDark,zhang2025demultiplexing}, particularly through physics-informed deep learning~\cite{wang2025deep,glosemeyer2025real}.
Novel strategies are crucial for routers that enable far more complex connections using FMFs than the linear link between Alice and Bob investigated in this paper. Significantly longer connections are subject to stronger environmental influences. These can be addressed through new fibers~\cite{wong2012excitation} and intelligent optoelectronics, thereby allowing for scaling of the key rate by utilizing a larger number of spatial channels.





\bibliography{sample}

\section*{Acknowledgements}


This research was supported by the German Research Foundation for funding the Reinhart Koselleck project (project number: 560574412, CZ 55/61-1) and by the Federal Ministry of Education and Research of Germany with the project 6G-life (grant identification number: 16KISK001K) and QUIET (project identification number: 16KISQ092). We sincerely appreciate
funding by the Deutsche Forschungsgemeinschaft (Project
No. 552245798). Authors, Stefan Krause and Felix Kunzmann acknowledge the received funding from the European Regional Development Fund (ERDF) and from the state budget passed by the Saxon state parliament, reference numbers 100498890, 100498868 and 4-7324/27/3-2022/32856.
Thanks are going to Prof. Kay-Uwe Giering (Fraunhofer Institute for Integrated Circuits Dresden) for fruitful discussions on fiber-optical quantum communication. We also would like to thank Dr. Stefan Rothe for co-initiating the ideas for combining deep learning and quantum for demultiplexing strategies.

\section*{Author contributions statement}

Q.Z., J.C. and S.K. conceived the concept. 
Q.Z. and S.K. conceived the experiments, S.K., Q.Z. and F.K. conducted the experiments, S.K. and Q.Z. analysed the results. 
J.C. supervised the project.
All authors reviewed the manuscript.

\end{document}